\documentclass[11pt]{article}
\usepackage{moriond,epsfig}

\begin{document}
\rightline{LPT--Orsay 08/47}
\vspace*{4cm}
\title{WIMP MASS FROM DIRECT, INDIRECT DARK MATTER DETECTION EXPERIMENTS AND COLLIDERS: A COMPLEMENTARY AND MODEL-INDEPENDENT APPROACH.}

\author{ Nicol\'as Bernal }

\address{Laboratoire de Physique Th\'eorique, B\^atiment 210,\\
Universit\'e Paris-Sud XI, 91405 Orsay Cedex, France}

\maketitle\abstracts{
We study the possibility of identifying dark matter properties from direct (XENON$100$) and indirect (GLAST) detection experiments.
In the same way, we examine the perspectives given by the next generation of colliders (ILC).
All this analysis is done following a model-independent approach.
We have shown that the three detection techniques can act in a highly complementary way.
whereas direct detection experiments will probe efficiently
light WIMPs, given a positive detection (at the $10\%$ level for
$m_{\chi} \sim 50$ GeV), GLAST will be able to confirm
and even increase the precision in the case of a NFW profile, for a
WIMP-nucleon cross-section $\sigma_{\chi-p}<10^{-8}$ pb.
However, for heavier WIMP ($\sim 175$ GeV), the ILC will lead the reconstruction of the mass.}

\section{Direct detection}
Dark Matter (DM) direct detection experiments measure the elastic collisions between WIMPs and target nuclei in a detector, as a function of the recoil energy $E_r$.
The detection rate depends on the density $\rho_0\simeq 0.3$ GeV cm$^{-3}$ and velocity distribution $f(v_\chi)$ of WIMPs near the Earth (a Maxwellian halo will be considered).
The differential rate per unit detector mass and per unit of time can be written as:
\begin{equation}
\frac{dN}{dE_r}=\frac{\sigma_{\chi-N}\,\rho_0}{2\,m_r^2\,m_\chi}\,
F(E_r)^2\int_{v_{min}(E_r)}^{\infty}\frac{f(v_\chi)}{v_\chi}\,dv_\chi\, ,
\label{Recoil}
\end{equation}
where $\sigma_{\chi-N}$ is the WIMP-nucleus scattering cross section, $m_\chi$ the WIMP mass and $m_r$ is the WIMP-nucleon reduced mass.
$F(E_r)$ is the nucleus form factor; we assume it has the Woods-Saxon form.

The XENON \cite{Angle:2007uj} experiment aims at the direct detection of DM via its elastic scattering off xenon nuclei.
In this study we will consider the case of a $100$ kg Xenon-like experiment and $3$ years taking data.
We consider $7$ energy bins between $4$ and $30$ keV.
For this experiment we took a zero background scenario; of course a more detailed analysis could take into account non-zero background simulating the detector and in particular the neutron spectrum.
So, in that sense our results will be optimistic.

One option to discriminate between a DM signal and the background is to use the $\chi^2$ method.
Let us call $N^{sign}$ the signal, $N^{bkg}$ the background and $N^{tot}\equiv N^{sign}+N^{bkg}$ the total signal measured by the detector.
For an energy range divided into $n$ bins the $\chi^2$ is defined as:
\begin{equation}
\chi^2=\sum^n_{i=1}\left(\frac{N_i^{tot}-N_i^{bkg}}{\sigma_i}\right)^2.
\end{equation}
We assume a Gaussian error $\sigma_i\equiv\sqrt{\frac{N_i^{tot}}{M\cdot T}}$ on the measurement, where $M$ is the detector mass and $T$ the exposure time.

\section{Indirect detection}
The spectrum of gamma--rays generated in dark matter annihilations
and coming from the galactic center can be written as
\begin{equation}
\Phi_\gamma(E_\gamma)=\sum_i\frac{dN_{\gamma}^i}{dE_{\gamma}}\,Br_i\,\langle\sigma v\rangle\,\frac{1}{8\pi\,m_{\chi}^2}\int_{line\ of\ sight} \rho^2\ dl\ ,
\end{equation}
where the discrete sum is over all dark matter annihilation
channels,
$dN_{\gamma}^i/dE_{\gamma}$ is the differential gamma--ray yield,
$\langle\sigma v\rangle$ is the annihilation cross-section averaged
over the WIMPs' relative velocity distribution and $Br_i$ is the branching ratio of annihilation into the \textit{i}-th final state.
It is possible to concentrate ourselves on a process which gives $100\%$ annihilation into $WW$ pairs, as this choice will not influence significantly the result of the study \cite{Bernal:2008zk}.
The dark matter density $\rho$ is usually parametrized as
\begin{equation}
\rho(r)=\frac{\rho_0}{(r/R)^{\gamma}\,[1+(r/R)^{\alpha}]^{(\beta-\gamma)/\alpha}}.
\end{equation}
We assume a NFW profile with $\alpha=\gamma=1$, $\beta=3$ and $R=20$ kpc, producing a profile with a behavior $\rho(r)\propto r^{-\gamma}$ in the inner region of the galaxy.

The gamma-ray telescope GLAST \cite{Gehrels:1999ri} will perform an all-sky survey covering an energy range $1 - 300$ GeV.
We will consider a GLAST-like experience with an effective area and angular resolution on the order of $10^4$ cm$^2$ and $0.1^\circ\times0.1^\circ$ ($\Delta\Omega\simeq 10^{-5}$ sr) respectively, who will be able to point and analyze the inner centre of our galaxy.
We consider also $3$ years of effective data acquisition experiment.\\
For this experiment, the background can be modeled by interpolating \cite{Bernal:2008zk} the gamma-ray spectrum measured by HESS \cite{Aharonian:2004wa} (for $E_\gamma>160$ GeV) and EGRET \cite{Hunger:1997we} (for $E_\gamma<10$ GeV) missions.

\section{Colliders}
Recently an approach was proposed by A. Birkedal et al.\cite{Birkedal:2004xn} which allows to perform a model-independent study of WIMP properties at lepton colliders.
Since the known abundance of DM gives specific values for the DM annihilation cross section, one might hope this cross section can be translated into a rate for a measurable process at a collider.\\
The starting point is to relate total annihilation cross section to the cross section into $e^+e^-$ pairs
\begin{equation}
\kappa_e\equiv\sigma(\chi\chi\rightarrow e^+e^-)/\sigma(\chi\chi\rightarrow all).
\end{equation}
Then we can use the \textit{detailed balancing} equation to relate $\sigma(\chi\chi\rightarrow e^+e^-)$ to $\sigma(e^+e^-\rightarrow\chi\chi)$, for non-relativistic WIMPs.
But this kind of process containing only WIMPs in the final state is not visible in a collider since they manifest themselves just as missing energy.
However this process can be correlated to the radiative WIMP pair-production $\sigma(e^+e^-\rightarrow\gamma\,\chi\chi)$ using the \textit{collinear factorization}.
This approach is valid for photons which are either soft or collinear with respect to the colliding beams.
The accuracy of the approximation outside the previous region has been discussed \cite{Birkedal:2004xn} with the conclusion that the approach works quite well.\\
So, starting from the total annihilation cross section $\sigma_{an}$ we can compute $\sigma(e^+e^-\rightarrow\gamma\,\chi\chi)$
\begin{equation}
\frac{d\sigma(e^+e^- \rightarrow \chi\chi\gamma)}{dx\,dcos\theta} \approx \frac{\alpha\,\kappa_e\,\sigma_{an}}{16\,\pi}\,\frac{1+(1-x)^2}{x}\,\frac{1}{\sin^2\theta}\,2^{2\,J_0}\,(2S_\chi +1)^2\,\left(1-\frac{4\,m_\chi^2}{(1-x)\,s}\right)^{1/2+J_0}\, .
\end{equation}
Here $x\equiv 2\,E_\gamma/\sqrt{s}$, $\theta$ is the angle between the photon and the incoming beam, $S_\chi$ and $J_0$ are the spin of the WIMP and the dominant value of the angular momentum in the velocity expansion for $\langle\sigma v\rangle$.\\
We place ourselves in the framework of the ILC project with a center-of-mass energy of $\sqrt{s}=500$ GeV and an integrated luminosity of $500$ fb$^{-1}$.
For this process with only a single photon detected, the main background in the standard model is radiative neutrino production \cite{Bernal:2008zk}.

\section{Complementarity}
\begin{figure}
\begin{center}
\includegraphics[angle=270,scale=0.31]{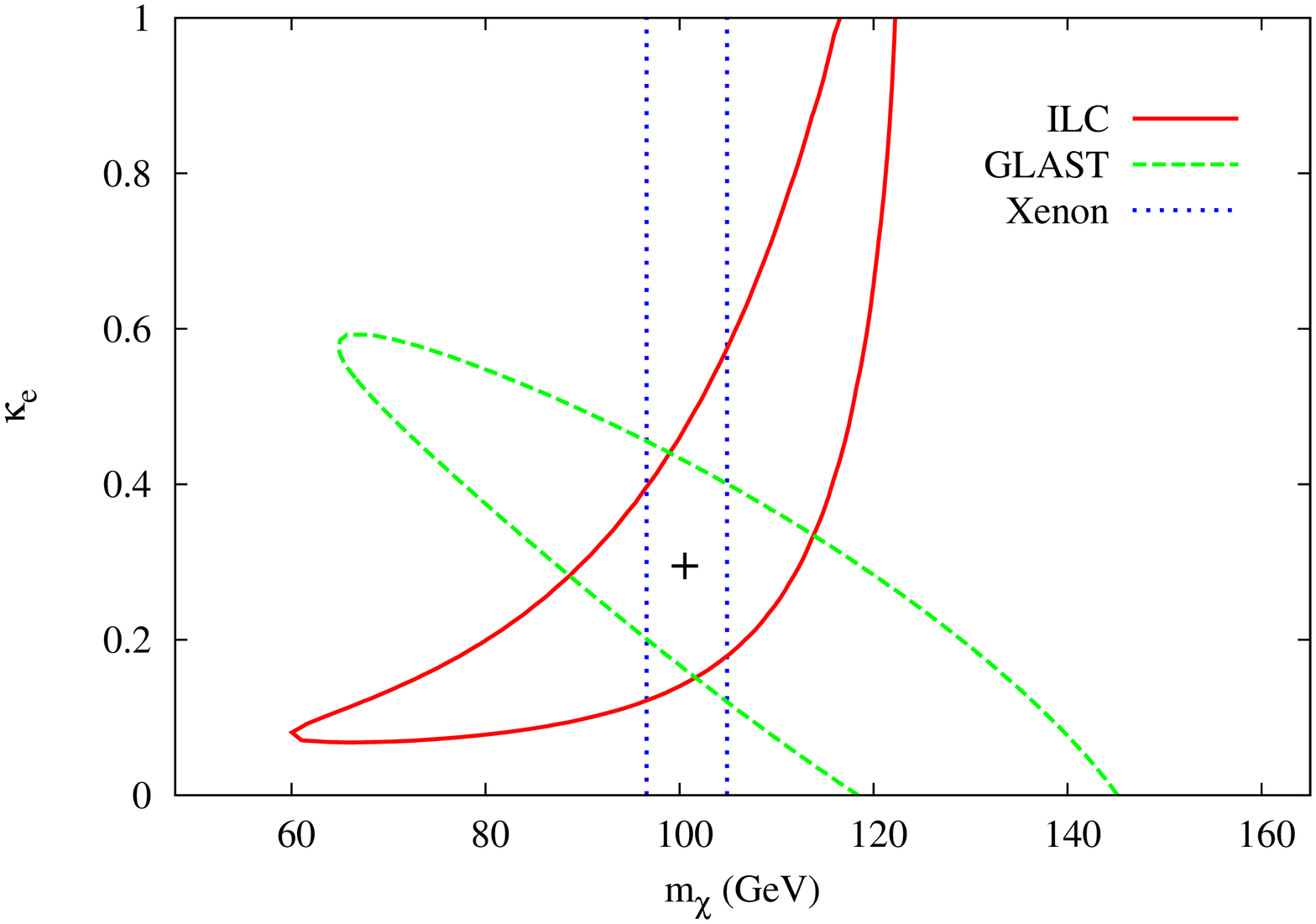}
\includegraphics[angle=270,scale=0.31]{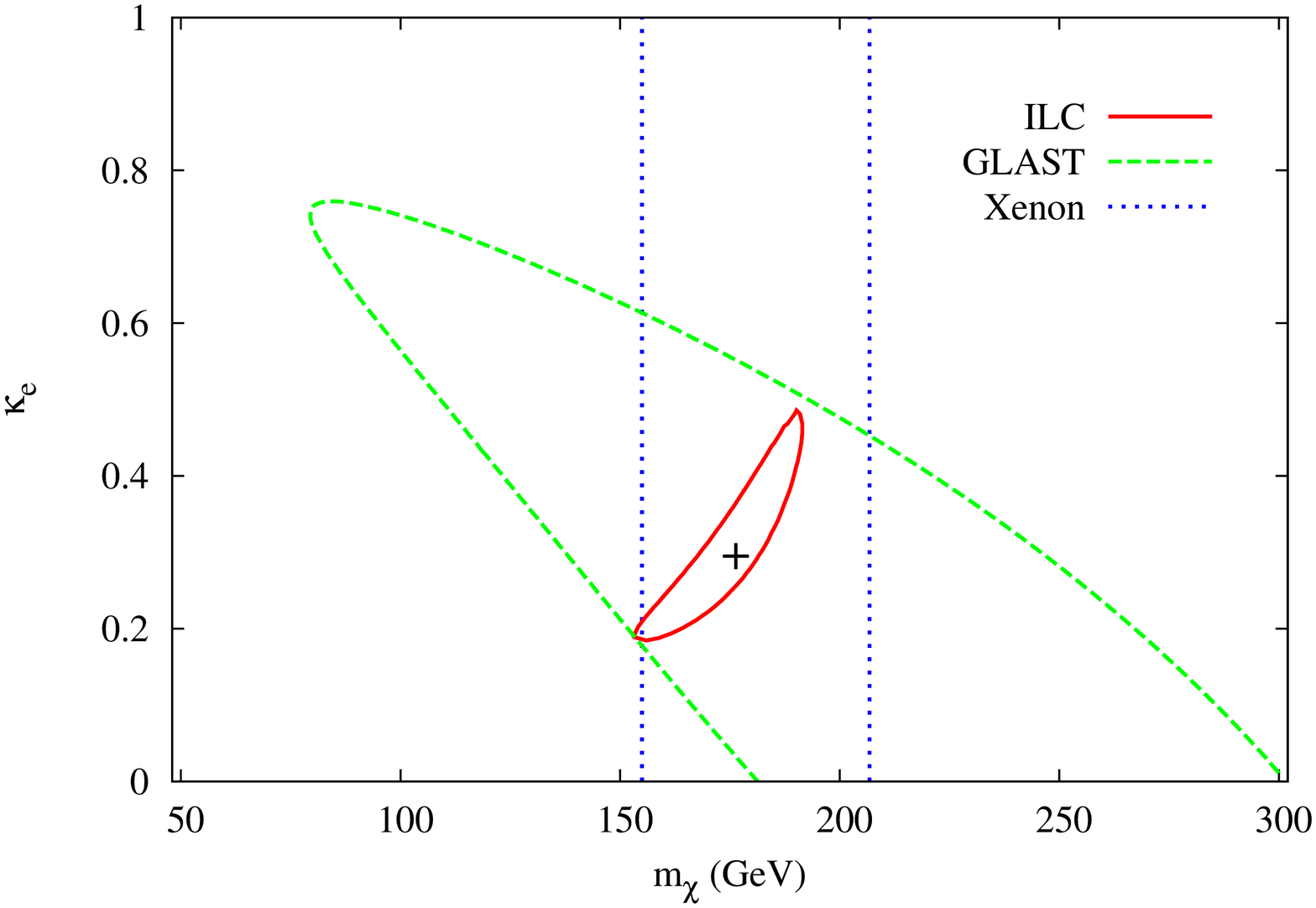}
\caption{{\footnotesize
Comparison between a $100$ kg XENON-like experiments (dotted line) with
$\sigma_{\chi-p}=10^{-7}$ pb, GLAST (dashed line) in the case of an NFW halo
 profile with $\langle\sigma v\rangle=3\cdot10^{-26}$ cm$^3$s$^{-1}$, and
 unpolarized ILC sensitivity (solid line) at $2\sigma$ of confidence level, for
 different WIMP masses $m_\chi=100$ and $175$ GeV, and $\kappa_e=0.3$.}}
\label{figura}
\end{center}
\end{figure}

Recently, several works have studied the determination of the WIMP mass for the case of direct\cite{Green} and indirect \cite{Dodelson:2007gd} detection experiments.
Furthermore, Drees and Shan \cite{Drees} showed that one can increase such a precision with a combined analysis of two experiments of direct detection.\\
In figure \ref{figura} we compare the precision levels for direct and indirect detection experiments, along with the corresponding results of the method we followed for the ILC for $\kappa_e=0.3$ and two cases of WIMP masses $m_\chi=100$ (left panel) and $175$ GeV (right panel).
All the results are plotted for a $2\sigma$ confidence level.
The green-dashed lines correspond to the results for a GLAST-like experiment assuming a NFW halo profile.
The total annihilation cross-section has been taken to be $\langle\sigma v\rangle=3\cdot 10^{-26}$ cm$^3$s$^{-1}$.
The red-plain line represents the result for an ILC-like collider with non-polarized beams.
The blue-dotted line corresponds to a $100$ kg XENON-like experiment, assuming a WIMP-nucleus scattering cross-section of $10^{-7}$ pb.
All the parameter space points that lie within the marked regions can not be discriminated by the corresponding experiments.\\
It is pertinent to study the complementarity between the three experiences listed above firstly because the mass reconstruction yields comparable results, hence a combination of these data can substantially improve the final result.
Secondly, because we can probe different regions in the parameter space.\\
For the case of a $100$ GeV WIMP, a GLAST- or an ILC-like experiment alone can provide a limited precision for the WIMP mass ($\sim 60\%$).
Combined measurements can dramatically increase the precision , reaching an accuracy of $\sim 25\%$.
If we additionally include direct detection measurement, we can reach a precision of the order of $\sim 9\%$.

In table \ref{tabla} we show the precision expected for several dark DM masses.
A light WIMP mass ($\sim 50$ GeV) can be reconstructed by both direct and indirect DM experiments with a high level of precision; however for the ILC the model independent procedure fails because of the relativistic nature of the WIMP.
On the contrary, the ILC will be particularly efficient to measure a WIMP with a mass of about $175$ GeV.
Concerning a $500$ GeV WIMP, only a loose lower bound could be extracted form direct and indirect experiments.
In this case the ILC will not be kinematically able to produce so heavy WIMPs.
\begin{table}[t]
\caption{{\footnotesize Precision on a WIMP mass expected from the
different experiments at $2\sigma$ after $3$ years of exposure,  
for $\sigma_{\chi-p}=10^{-7}$ pb, a NFW profile and a 
$500$ GeV unpolarized linear collider with a luminosity of $500$ fb$^{-1}$}.
All values are given in GeV.}
\vspace{0.4cm}
\begin{center}
\begin{tabular}{|c|c|c|c|}
\hline 
$m_{\chi}$ & XENON & GLAST & ILC   \\
\hline 
$50$  & $\pm$ $1$ & $\pm$ $8$ & $**$ \\
$100$ & $\pm$ $6$ & $-25/+32$ & $-40/+20$ \\
$175$ & $-25/+35$  & $-70/+100$ & $-20/+15$ \\
$500$ & $-250/**$  & $-350/**$  & $**$ \\
\hline 
\end{tabular}
\label{tabla}
\end{center}
\end{table}

\section*{Acknowledgments}
I would like to thank the organising committee for inviting me at this pleasant conference.
Likewise, I would like to thank the ENTApP Network of the ILIAS project RII3-CT-2004-506222 and the French ANR project PHYS@COL\&COS for financial support.
The work reported here is based in collaboration with A. Goudelis, Y. Mambrini and C. Mu\~noz.


\section*{References}
\begin{thebibliography}{99}
\bibitem{Angle:2007uj}
  J.~Angle {\it et al.}  [XENON Collaboration],
  %``First Results from the XENON10 Dark Matter Experiment at the Gran Sasso
  %National Laboratory,''
  Phys.\ Rev.\ Lett.\  {\bf 100} (2008) 021303
  [arXiv:0706.0039 [astro-ph]].
  %%CITATION = PRLTA,100,021303;%%
\bibitem{Bernal:2008zk}
  N.~Bernal, A.~Goudelis, Y.~Mambrini and C.~Mu\~noz,
  %``Determining the WIMP mass using the complementarity between direct and
  %indirect searches and the ILC,''
  arXiv:0804.1976 [hep-ph].
  %%CITATION = ARXIV:0804.1976;%%
\bibitem{Gehrels:1999ri}
  N.~Gehrels and P.~Michelson,
  %``GLAST: The next-generation high energy gamma-ray astronomy mission,''
  Astropart.\ Phys.\  {\bf 11}, 277 (1999).
  %%CITATION = APHYE,11,277;%%
\bibitem{Aharonian:2004wa}
  F.~Aharonian {\it et al.}  [The HESS Collaboration],
  %``Very high energy gamma rays from the direction of Sagittarius A*,''
  Astron.\ Astrophys.\  {\bf 425} (2004) L13
  [arXiv:astro-ph/0408145].
  %%CITATION = AAEJA,425,L13;%%
\bibitem{Hunger:1997we}
  S.~D.~Hunger {\it et al.},
  %``EGRET observations of the diffuse gamma-ray emission from the galactic
  %plane,''
  Astrophys.\ J.\  {\bf 481} (1997) 205.
  %%CITATION = ASJOA,481,205;%%
\bibitem{Birkedal:2004xn}
  A.~Birkedal, K.~Matchev and M.~Perelstein,
  %``Dark matter at colliders: A model-independent approach,''
  Phys.\ Rev.\  D {\bf 70}, 077701 (2004)
  [arXiv:hep-ph/0403004].
  %%CITATION = PHRVA,D70,077701;%%
\bibitem{Green}
  A.~M.~Green,
  %``Determining the WIMP mass using direct detection experiments,''
  JCAP {\bf 0708} (2007) 022
  [arXiv:hep-ph/0703217];
  %%CITATION = JCAPA,0708,022;%%
  A.~M.~Green,
  %``Determining the WIMP mass from a single direct detection experiment, a more
  %detailed study,''
  arXiv:0805.1704 [hep-ph].
  %%CITATION = ARXIV:0805.1704;%%
\bibitem{Dodelson:2007gd}
  S.~Dodelson, D.~Hooper and P.~D.~Serpico,
  %``Extracting the Gamma Ray Signal from Dark Matter Annihilation in the
  %Galactic Center Region,''
  Phys.\ Rev.\  D {\bf 77} (2008) 063512
  [arXiv:0711.4621 [astro-ph]].
  %%CITATION = PHRVA,D77,063512;%%
\bibitem{Drees}
  M.~Drees and C.~L.~Shan,
  %``Model-Independent Determination of the WIMP Mass from Direct Dark Matter
  %Detection Data,''
  arXiv:0803.4477 [hep-ph];
  %%CITATION = ARXIV:0803.4477;%%
  C.~L.~Shan and M.~Drees,
  %``Determining the WIMP Mass from Direct Dark Matter Detection Data,''
  arXiv:0710.4296 [hep-ph].
  %%CITATION = ARXIV:0710.4296;%%

\end{thebibliography}
\end{document}